\documentclass[a4paper]{jpconf}

\usepackage{graphicx}

\usepackage{hyperref}

\usepackage[sort&compress,numbers]{natbib}

\usepackage{amsmath}
\usepackage{amssymb}
\usepackage{bm}

\newcommand{\castro}{{\sffamily Castro}}

\newcommand{\amrex}{{\sffamily AMReX}}
\newcommand{\vode}{{\sffamily VODE}}

\newcommand{\Uc}{{\,\bm{\mathcal{U}}}}
\newcommand{\Adv}[1]{{\left [\boldsymbol{\mathcal{A}} \left(#1\right)\right]}}

\newcommand{\isot}[2]{$^{#2}\mathrm{#1}$}

\newcommand{\gcc}{\mathrm{g~cm^{-3} }}

\usepackage{color}
\begin{document}

\title{Toward Resolved Simulations of Burning Fronts in Thermonuclear
       X-ray Bursts}

\author{M. Zingale$^1$,
        K.~Eiden$^1$,
        Y.~Cavecchi$^{2,3}$,
        A.~Harpole$^1$,
        J.~B. Bell$^4$,
        M.~Chang$^1$,
        I.~Hawke$^3$,
        M.~P. Katz$^5$,
        C.~M. Malone$^6$,
        A.~J. Nonaka$^4$,
        D.~E. Willcox$^4$, and
        W. Zhang$^4$}

\address{$^1$Department of Physics and Astronomy, Stony Brook
  University, Stony Brook, NY 11794-3800 USA}

\address{$^2$Department of Astrophysical Sciences, Princeton University,
  Peyton Hall, Princeton, NJ 08544, USA}

\address{$^3$Mathematical Sciences and STAG Research Centre,
  University of Southampton, SO17 1BJ, UK}

\address{$^4$Center for Computational Sciences and Engineering,
  Lawrence Berkeley National Lab, Berkeley, CA 94720 USA}

\address{$^5$NVIDIA Corporation, 2788 San Tomas Expressway,
  Santa Clara, CA, 95050 USA}

\address{$^6$Los Alamos National Laboratory, Los Alamos, NM, 87545 USA}

\ead{michael.zingale@stonybrook.edu}

\begin{abstract}
We discuss the challenges of modeling X-ray bursts in
multi-dimensions, review the different calculations done to date, and
discuss our new set of ongoing simulations.  We also describe
algorithmic improvements that may help in the future to offset some of
the expense of these simulations, and describe what may be possible
with exascale computing.
\end{abstract}

\section{Introduction}

X-ray bursts (XRBs) are fascinating astrophysical explosions.  A
neutron star accretes fuel from its binary companion, building up just
a thin layer ($\sim 5$--$10$~m) before the immense gravitational
acceleration at the neutron star surface compresses and heats the fuel
to the point of thermonuclear runaway.  A brief flash of X-rays
follows, the accreted layer diffuses the heat produced from reactions,
and the process repeats.  We can observe multiple bursts from a single
source, and the information their lightcurves encodes tells us about
the underlying neutron star, and ultimately the nuclear equation of
state.  A challenging aspect of interpreting the lightcurves is
understanding the radiation transport through the hot ash layers.  In
particular, the composition can affect the effective temperature,
which in turn affects the radius we measure for the neutron
star~\cite{suleimanov:2011}.  Convection in the burning layer can even
bring ashes up to the photosphere which alters what we
see~\cite{kajava:2017}.  Accurate measurements of neutron star radii
from XRBs could greatly constrain the nuclear equation of
state~\cite{steiner:2010,ozel:2010}.  Finally, brightness oscillations
during the rising phase of the bursts
and other features seen in lightcurves are interpreted as arising from
the spreading of a hotspot across the neutron star---demonstrating
that the burst begins in a localized
region~\cite{bhattacharyya:2006,bhattacharyya:2007}.
See~\cite{galloway:2017} for a recent review of XRBs.

Simulations of XRBs can help us to understand the observations, but
many technical challenges remain.  The primary difficulties are simply
the range of length and timescales involved.  At the largest lengthscale,
the neutron star radius is 10~km and the Rossby lengthscale, where rotation
and lateral pressure gradients balance, is $\sim
1$~km~\cite{SPIT_ETAL02}.  At the smallest scales, a conductive He
flame has a thickness of 10s of centimeters~\cite{Timmes00} and the
pressure scale height of the fuel layer is on the order of meters.  To capture the
rise timescale of the XRB, we would need to model seconds, but the flame itself
is subsonic, so long timescale evolution is difficult.

Present and past simulations of XRBs employ a variety of
approximations to get past these difficulties.  The variety of
approximations used enables a complementary picture of XRBs to be built up,
and has greatly advanced our understanding of these systems.
The breadth of simulations to date has been impressive.  We summarize
the different approaches below.

One-dimensional simulations, e.g. \cite{woosley-xrb,fisker:2008}, assume
spherical symmetry and model the burst through the vertical column
depth of fuel.  These simulations can capture the luminosities and
burst recurrence times well.  They can utilize very large nuclear
reaction networks and have been used to understand the nucleosynthesis
and rp-process~\cite{schatz:rp1999,rpprocess,schatz_rp}, as well as to understand how rate uncertainties can
affect the outcomes of the bursts.  Many successive bursts can be
modeled and these simulations can show us the evolution of the ash
layer with time.  The approximation of 1D however means that we cannot
learn about lateral variations across the neutron star, including
flame spreading.

Global shallow water simulations~\cite{SPIT_ETAL02} capture the
large scale spreading of the burning front by using a very simple
vertical structure.  These simulations were instrumental in showing
that the Coriolis force plays an important role in confining a
spreading region and that the flame spreading can explain brightness
oscillations during the rise of the burst lightcurve.

Our group and others have modeled the convection preceding the
runaway using low Mach number hydrodynamic methods for pure helium
bursts~\cite{Lin:2006,xrb} and mixed hydrogen/helium
bursts~\cite{xrb2} in 2D, as well as full 3D simulations of mixed
bursts~\cite{xrb3}, and followed the development of turbulence.
These simulations can give an understanding of the pre-burning front
regime, including the strength and character of the turbulence, but
cannot model large scale lateral variations because of the implicit
assumptions built into the low Mach number model~\cite{ABRZ:I}.

While detonations are computationally easier to model than flames (we
largely get the speed through the jump conditions in the Riemann
problem without resolving the structure), detonations require extreme
conditions not found in XRBs~\cite{ZINGALE_ETAL01,harpole:2018}.
This means that calculations that want to capture the nucleosynthesis
yields from spreading burning fronts in XRBs need to model a deflagration.

Flame spreading was first modeled in a series of calculations
employing an algorithm from atmospheric science, where vertical
hydrostatic equilibrium is enforced and wide-aspect ratio zones give a
horizontal CFL number that is large enough to obtain long timescale
evolution~\cite{cavecchi:2012}.  These calculations display key
features of the propagation mechanism, namely a balance between
hydrodynamics and the pure flame physics. The flame proceeds mainly
via conduction, thus being a deflagration. The flame front, however,
is not vertical but inclined, the angle with the horizontal being very
small: $\theta \sim H/R_{\rm{Ro}} \sim 10^{-3}$. Here $H$ is the scale
height and $R_{\rm{Ro}}$ is the Rossby radius. It is the
hydrodynamical balance between the Coriolis force, gravity and
pressure gradients that determines the Rossby radius and therefore the
inclination angle of the flame front. The extended surface of the
flame front is what leads to propagation speeds of order $10^5 $~km/s,
in good agreement with observations. Due to its dependence on the
Coriolis force, the flame speed changes across the surface of the NS
(being faster at the equator), leading to potentially detectable
effects if the rise of the burst is observed with enough resolution
\cite{art-2015-cavecchi-etal}.  We note that there may be preexisting
turbulence ahead of the flame from simmering occurring
there, and no calculations have considered the effects of the flame
interacting with this turbulence, as we characterized in \cite{xrb3}.
Flame-turbulence interactions have long been seen as a means to
accelerate a flame in Type Ia supernovae, so future studies should
look at turbulence in the layer.  Capturing turbulence, of course,
requires high resolution, so the ideas described here will be needed.

The NS systems exhibiting XRBs are
expected to have magnetic fields in the range $10^7$--$10^{10}$~G.
Inclusion of a vertical magnetic field proved to have a strong
influence on the nature and the speed of flame propagation
\cite{art-2016-cavecchi-etal}, due to the interaction between the
Coriolis force and the magnetic tension, which further acts towards
determining the inclination angle of the flame front.

A longstanding goal is to perform simulations where we resolve the
flame structure, to allow us to accurately capture the
nucleosynthesis, and watch it move across the neutron star surface.
The results of these simulations will enable strict comparison to
observations made with the new generation of X-ray telescopes such as
\textit{eXTP}\/ and \textit{STROBE-X}\/
\cite{art-2017-wilhod-etal,art-2016-zhang-etal,intZand2018} which will offer
high collecting area and time resolution.
We show some in-progress calculations that are a step toward this, discuss
their remaining approximations, and also discuss what new algorithmic
techniques might be needed to realize this goal with the advent of
exascale computing.

\section{Resolved Flame Studies}

We have begun a set of flame spreading calculations using the
compressible hydrodynamics code \castro~\cite{castro}, part of the
AMReX astrophysics suite~\cite{astronum:2017}.  \castro\ uses adaptive
mesh refinement, supports a general equation of state and reaction
networks, has a conservative gravity formulation~\cite{wdmergerI}, and
is optimized to run on current supercomputers.  All of our code base
is open source and available on
GitHub\footnote{\url{http://github.com/amrex-astro/}}, and all of the
solvers, problem setup, initial models, and analysis scripts for each
of our science problems are distributed with the codes.

We setup a problem by creating a pair of 1D hydrostatic XRB models (hot and cool)
representing the post- and pre-flame states and apply these laterally
across the domain to setup a hot region at the left end of our domain
in equilibrium with the cool model on the right.  Thermal diffusion and nuclear
reactions quickly produce a laterally propagating flame that begins to
move through our domain.  We resolve the diffusion scale, using a
resolution of 10~cm, and use adaptive mesh refinement (two jumps of
$4\times$) to keep a buffer of low density material at low resolution
above the surface to allow for expansion.  To make the lengthscales
tractable, we rotate the neutron star at 2000~Hz---much faster than
expected.  This helps confine the spreading region to a more
reasonable horizontal domain.  Nevertheless, our simulation uses 12288
zones laterally, on the finest grid.  We use the stellar
conductivities from \cite{Timmes00} and a 13-isotopic alpha-chain
reaction network.  To make the timescales easier, we boost the flame
speed by a factor of 10 by scaling both the conductivity and energy
generation rate by the same factor.  This is an approximation we hope to
relax in the near future.

\begin{figure}[t]
\centering
\includegraphics[width=\linewidth]{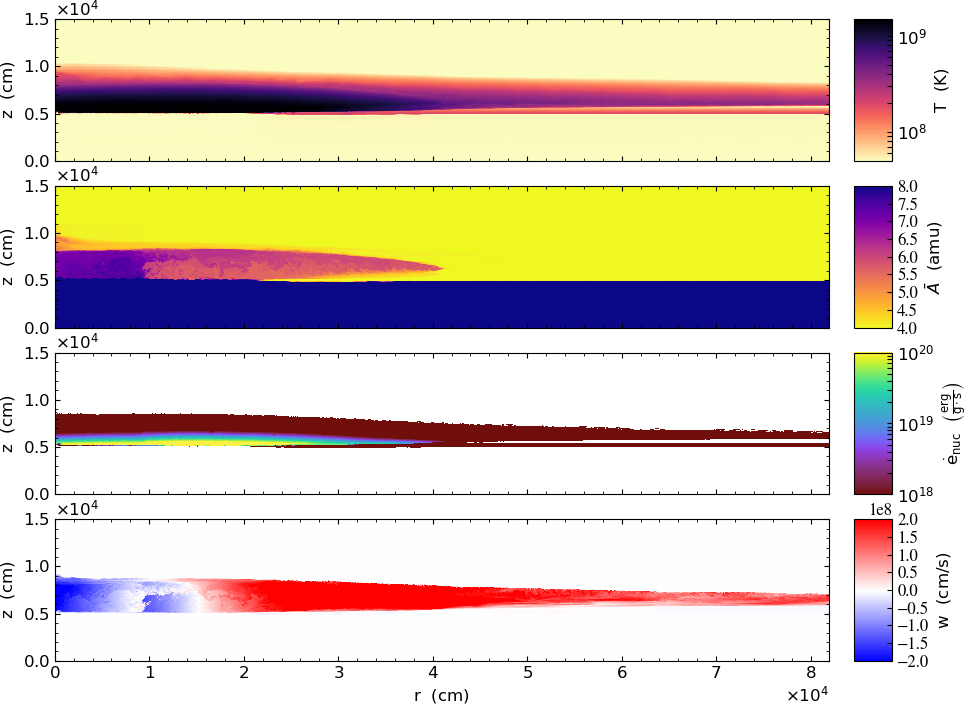}
\caption{\label{fig:xrb} Helium flame spreading across the surface of
  a neutron star.  Shown are temperature (top), the mean molecular
  weight of the ash (second), the nuclear energy generation rate
  (third), and the velocity out of the simulation plane (bottom).  The
  out-of-plane velocity contours indicate where the Coriolis force has
  created a hurricane-like wind structure which confines the flame,
  leading to an inclined front. It is at such an interface that the
  temperature and burning rate are highest.  }
\end{figure}

\begin{figure}[t]
\centering
\includegraphics[width=0.9\linewidth]{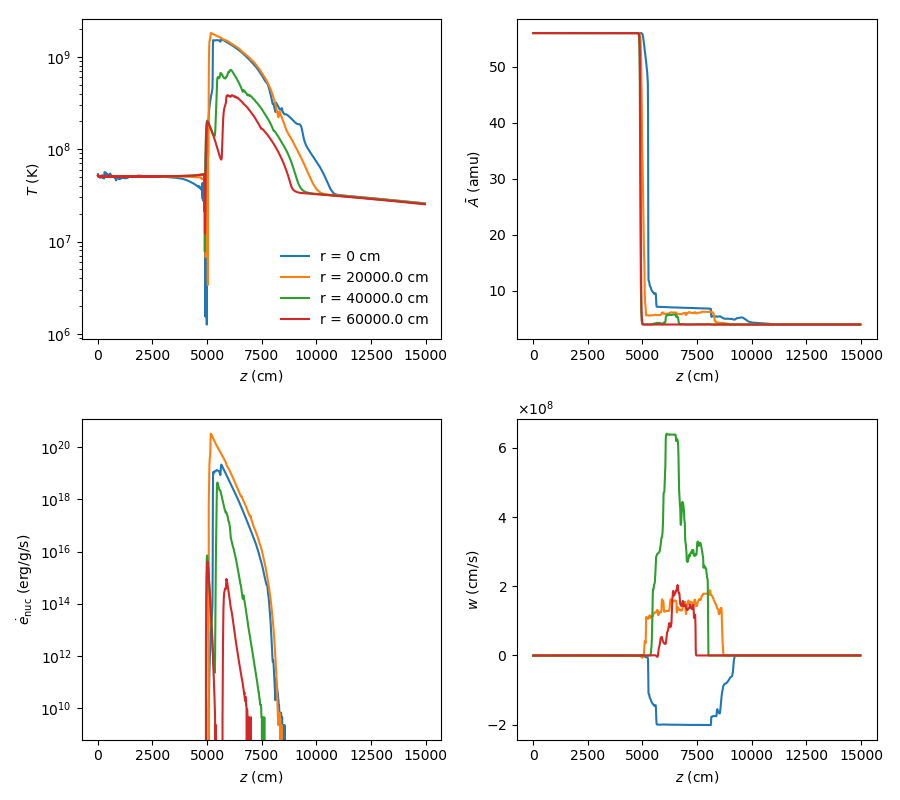}
\caption{\label{fig:xrb_profiles} Vertical profiles at several radial distances
  from the vertical axis through the plots shown in Figure~\ref{fig:xrb}.  These
  profiles sample the region behind and ahead of the flame.}
\end{figure}

Figure~\ref{fig:xrb} shows the flame structure for an initial 2D run
at 0.007~s.  Vertical profiles at a few radial distances from the
origin are shown for the same quantities in
Figure~\ref{fig:xrb_profiles}.  The burning layer is only about 10~m
high.  Underneath is \isot{Ni}{56}, representative of the underlying
neutron star, and above it is a very low density ($10^{-4}~\gcc$)
buffer.  Only a portion of the vertical extent is shown.  The top
panel shows the temperature structure.  The initial perturbation
reached to $2\times 10^4~\mathrm{cm}$, so at this point, the extent of
the burnt region has nearly doubled.  The flame structure is quite
wide, since there are multiple burning stages beyond just helium to
carbon, and this shows up in the composition.  This is seen more
clearly in the next panel which shows the mean molecular weight of the
nuclei ($\bar{A}$).  We see that the further behind the head of the
flame we are, the heavier the ash nuclei.  We also see the head of the
flame appears lifted off of the base of the burning layer---unburned
material underlies the spreading flame.  This is reminiscent of the
flame structure seen in the earlier simulations of
\cite{cavecchi:2012}.  The third panel shows the energy generation
rate.  It is strongest just above the neutron star surface, behind the
flame.  We again see that the head of the flame is detached from the
base of the atmosphere, with the strongest burning at the head of the
flame slightly higher up in the atmosphere.  Finally, the bottom panel
shows the velocity through the plane of the simulation---this is
induced by the Coriolis force as the burning fluid expands and begins
to spread laterally.  A tight hurricane structure has setup as a
result of the flame spreading.  This is the geostrophic balance
discussed in \cite{SPIT_ETAL02}.  These calculations are ongoing and
will be the subject of a detailed study in the near future.

\section{Large Scale Simulations and Algorithmic Developments}

The 2D calculations shown in the previous section require about 10k
node-hours of computational time to get to about 10~ms (running
on the Edison machine at NERSC).  We expect qualitatively different
results in 3D calculations: for example, the shear from the hurricane structure
acting along the flame front may cause instabilities that can affect the flame
propagation.
With our current simulation methodology,
a 3D version of this calculation is estimated to require 10M node-hours
or more---this is about the size of an annual allocation.
This simulation covers less than
1\% of the neutron star surface, so a fully resolved calculation of
burning over the entire neutron star is out of reach: a naive scaling
calculation suggests that 15 years of Moore's Law improvements are required.
%
%

Improvements are needed to push the scale of simulations
to the point where we can model XRBs at the full star scale with
resolved nuclear physics.
As the computational cost is roughly a product of the number of zones, the
number of zone updates, and the cost per zone update, we need to look for
improvements in each of these. The number of zones depends on the physical
model employed, the meshing strategy, and the accuracy required. The number of
zone updates depends on the smallest zone size and the largest characteristic
speed in the model. The computational cost per zone update is dominated by the
time spent in the reactions and equation of state. We discuss a few
possible improvements that can come both from implementing new algorithms and
from exploiting new hardware.


Machine architectures are evolving, with heterogeneous architectures
becoming more common.  Our recent development in \castro\ has focused
on moving the hydrodynamics to GPUs~\cite{astronum:2017}.  The XRB
problem is an ideal case for this, since hydrodynamics with constant
gravity (for plane-parallel domains) ports in a straightforward manner
to GPUs.  The current GPU version of the \castro\ hydrodynamics solver
is about $10\times$ faster using GPUs on a node than the entire node
of CPU cores (comparing 4 NVIDIA Volta GPUs to a dual-socket IBM Power9 CPU using 40 cores).
We see similar speed ups with reaction networks, with these GPU ports
under active development.  This means a GPU calculation could lower
the above costs by a factor of 10.  GPU offloading of the reaction
networks will also allow us to explore larger networks and more
detailed nuclear physics.

Another way we can reduce the number of zones for given accuracy is to use a more
accurate hydrodynamic method.  We are developing fully
fourth-order (in space and time) coupling of hydrodynamics and
reactions using the method of spectral deferred corrections (SDC) in
\castro, instead of the more commonly employed Strang splitting~\cite{strang:1968}.  This
provides us with two benefits.  First, the improved coupling actually
reduces the stiffness of the reactions, requiring fewer righthand side
calls and therefore reducing the expense of the
reactions.  Second, by moving to fourth
order, we may also be able to reduce our resolution requirements
needed for converged flames. For smooth wave solutions we may expect to need
around half the number of zones per dimension, and since computational work
scales like $(\Delta x)^4$ in 3D, this could lower costs by a factor of around
10.
%
%

\begin{figure}
\centering
\includegraphics[width=0.75\linewidth]{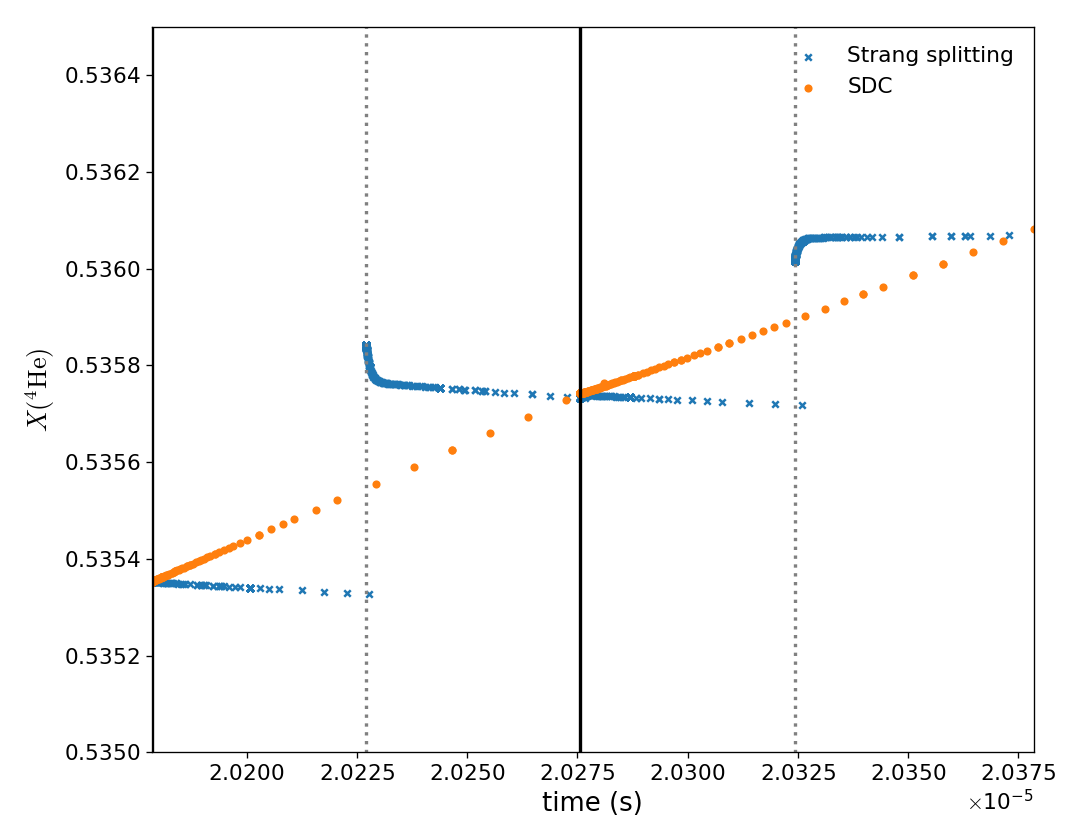}
\caption{\label{fig:sdc} Helium mass fraction as a function of time
  over two timesteps for a detonation test problem with a 19-isotope
  reaction network.  The points represent individual calls to the
  righthand side function of our reaction network (see
  Eq.~\ref{eq:syst}) and the density of points indicates how hard the
  integrator is working.  For the Strang case, we see that the first
  half of the reactions brings us off of the smooth solution
  (represented by the SDC curve), then the advection overcorrects,
  resulting in a discontinuous jump, and finally the second half of
  the reactions works to bring the state back to the solution curve.}
\end{figure}

As an example of the improvements in coupling between hydro
and reactions, Figure~\ref{fig:sdc} shows the mass fraction of helium
behind a detonation over the course of 2 timesteps.  For this test,
a 19 isotope reaction network was used.  The points
represent individual calls to the righthand side function of our
reaction network, and the density of
points indicates how hard the integrator is working.  We use the same
tolerances for both the Strang-split method and the SDC method.  For
the SDC method, we are using a second-order accurate method based on
\cite{SDC-old}.  We predict a time-centered advection term,
$\Adv{\Uc}^{n+1/2}_i$ using standard unsplit Godunov methods, but
explicitly include a reactive source term in the tracing of the
interface states, ${\bf R}(\Uc_i)$.  We can then solve the reactive
system, using this advective term as a piecewise-constant-in-time
source
\begin{equation}
\label{eq:syst}
\frac{d{\Uc_i}}{dt} = -\Adv{\Uc}^{n+1/2}_i + {\bf R}(\Uc_i)
\end{equation}
This system can be integrated with standard ODE methods.  To achieve
second-order accuracy, we need to iterate, using the updated $\Uc$ to
create a reactive source included in the predictor for the advective
update, and then reintegrate the system.  The result of this iteration
is that the advection sees the effects of reactions over the timestep
and the reactions see the effects of advection.  For the zone tracked
in Figure~\ref{fig:sdc}, each SDC iteration called the righthand side
function of the network 2 times less than the Strang case.  This
savings is dependent on the network and thermodynamic state, but our
tests thus far show that the SDC methods need fewer network calls to
integrate a timestep.  We also clearly see how the Strang system
evolves in a discontinuous fashion.  We are continuing to develop this
new integration strategy, with full fourth-order in space and time
reacting hydrodynamics methods to follow shortly.  The XRB simulations
are our target application.  We note that there has been some
development in the community on implicit time-integration methods for
compressible flow, but for the XRB problem, the wide range of Mach
number seen in the domain (exceeding unity near the top of the
atmosphere, ahead of the flame) removes much of the advantages of
implicit timestepping, so we focus on explicit methods.

For the ODE integrations shown we use the variable-order,
variable-step integrator \vode\ \cite{vode} with its implementation of
the backward-differentiation formulas.
We note that for both
Strang splitting and SDC, \vode\ must start the integration at first
order and correspondingly small timesteps since no higher-order
information about the solution is available (our reaction
network right-hand sides provide only the first time derivative of the
solution). Based on testing with a 13-isotope
approximate alpha-chain network, \vode\ takes many small timesteps
(e.g.\ $\sim 200$) before it can stably increase to a high order
representation of the solution, but after it does so, it can often
complete the integration in only a handful of steps (e.g.\ $\sim
5$--$6$). Observing that during a \castro\ timestep, the first Strang
step is a direct continuation of the integration in the second Strang
step of the previous \castro\ timestep, we explored saving the
\vode\ state following the second Strang step and restarting
\vode\ during the first Strang step. This approach allowed us to start
the first Strang step with a high order representation of the solution
and complete it in very few \vode\ timesteps ($\lesssim 10$ steps),
speeding up the \castro\ timestep advance by 50\% for our detonation
test problem using the 13-isotope network. However, this came at the
memory cost of storing $7 N_{eq} + 90$ scalar variables on our grid,
where for our networks $N_{eq} = N_{spec} + 2$ (in addition to the
number of species $N_{spec}$ we also integrate temperature and
specific energy). For a large-scale science problem with a large
network, this could quickly become prohibitive, so we are exploring
ways to mitigate either the memory requirements for restarting the
integration or the small timestep start up costs for the integration
method.

Other ways to make the simulation less expensive can include a subgrid
model. Flame models are commonly used for Type Ia supernovae.  There
however, the flame is thin compared to the scale height, so a simple
model can represent the unresolved flame on the grid.  For the XRB,
the flame thickness is a non-trivial fraction of the pressure scale
height and further, the laminar flame properties will change strongly
over the height of the burning layer (see~\cite{Timmes00}).  This
makes traditional flame models a poor fit.  Methods that use even
something as simple as subzone burning (e.g.~\cite{Wang2012190}) may
help capture the energetics with lower resolution.  This can be
implemented using the AMR machinery.  This may help a little, but will
still be insufficient for full star models.

If we continue to resolve the flame structure, then we need to be
smarter (and more aggressive) about our grid refinement strategy.
Currently, we put the entire fuel layer on the finest grid, but for
the material ahead of the flame, we could keep it at lower resolution
until the flame reaches it.  The main obstacle is that we need to keep
the fuel ahead of the flame in hydrostatic equilibrium, and keeping it
refined accomplishes this well.  Using well-balanced
schemes~\cite{kappeli:2016} can help us retain hydrostatic equilibrium
(HSE) in the lower refined regions, reducing the number of zones
needed to maintain fidelity, but we would still need a way to regrid
the layer ahead of the flame to higher resolution to accurately
capture the burning dynamics.  To accomplish this, we envision
building a refinement strategy that does interpolation of the data onto
the newly refined grids enforcing hydrostatic equilibrium.  This can
allow us to keep only a narrow fully refined region that ``slides''
along the surface of the neutron star keeping pace with the advancing
flame.  When using mesh refinement the overall costs are dominated by the
number of zones at the finest resolution, so such a scheme would greatly reduce
the computational costs. This is a longer term development.

All of the above developments still involve fully compressible
hydrodynamics, but multiscale methods~\cite{weinan2011principles}
where we couple different hydrodynamics solvers together in a single
simulation may ultimately be needed to model the full star while
resolving the burning front.
Such techniques rely on it being possible to decompose the dynamics of
the system into two scales, e.g.~a short and a long lengthscale, and
that the processes at each scale are at most weakly coupled with each
other. In the context of XRBs, one possibility for multiscale modeling
would be to model the large scale dynamics using a shallow water model
(like that of~\cite{SPIT_ETAL02}) to capture the effects of the
Coriolis force, and use a low Mach number or a fully compressible
model in the vicinity of the burning front to capture the turbulent
burning processes. Given that the computational costs of the shallow
water model would be negligible compared to those of the low
Mach/compressible model, the overall cost of this multiscale model
would be determined by the size of the low Mach/compressible region
only. This approach would therefore allow us to capture dynamics
across the full star without sacrificing resolution around the flame.
As with the more aggressive gridding strategy above, the hybrid
approach reduces the number of zones with the largest computational
cost through mesh refinement, and also reduces the computational costs
where possible by employing cheaper physical models in regions where
this is possible.  An example of coupling compressible and low Mach
number models implemented using the \amrex~framework, but in a
terrestrial context, is~\cite{Motheau2018}. An initial implementation,
\cite{Harpole2018}, in an astrophysical context, coupling shallow
water to compressible models in both Newtonian and relativistic
gravity, indicates that considerable computational efficiencies are
possible, but that close attention to the coupling between the models
is needed.  


\section{Summary}

XRBs are multiscale, multiphysics problems that are challenging to
model.  Nevertheless, significant progress has been made in
multidimensional models of these events, through a set of
complementary algorithmic approaches.  Future advances in computer
architectures and algorithms will allow for even more realistic models
of these events, and help us connect to observations, ultimately telling
us about the neutron star itself.

\ack The work at Stony Brook was supported by DOE/Office of Nuclear
Physics grant DE-FG02-87ER40317 and contract 7418390 with Lawrence
Berkeley National Laboratory as part of the Exascale Compute Project
ExaStar collaboration.  The work at LBNL was supported by the DOE
Office of Advanced Scientific Computing Research under Contract No,
DE-AC02-05CH11231. YC is supported by the European Union Horizon 2020
research and innovation programme under the Marie Sklodowska-Curie
Global Fellowship grant agreement No 703916.  An award of computer
time was provided by the Innovative and Novel Computational Impact on
Theory and Experiment (INCITE) program. This research used resources
of the Oak Ridge Leadership Computing Facility at the Oak Ridge
National Laboratory, which is supported by the Office of Science of
the U.S. Department of Energy under Contract No.\ DE-AC05-00OR22725.
This research used resources of the National Energy Research
Scientific Computing Center, which is supported by the Office of
Science of the U.S. Department of Energy under Contract
No.\ DE-AC02-05CH11231.  Visualizations were done using yt~\cite{yt}.
This research has made use of NASA's Astrophysics Data System
Bibliographic Services.


\bibliographystyle{iopart-num}
\bibliography{ws}

\providecommand{\newblock}{}
\begin{thebibliography}{10}
\expandafter\ifx\csname url\endcsname\relax
  \def\url#1{{\tt #1}}\fi
\expandafter\ifx\csname urlprefix\endcsname\relax\def\urlprefix{URL }\fi
\providecommand{\eprint}[2][]{\url{#2}}

\bibitem{suleimanov:2011}
{Suleimanov} V, {Poutanen} J and {Werner} K 2011 {\em \aap\/} {\bf 527} A139
  (\textit{Preprint} \eprint{1009.6147})

\bibitem{kajava:2017}
{Kajava} J~J~E, {N{\"a}ttil{\"a}} J, {Poutanen} J, {Cumming} A, {Suleimanov} V
  and {Kuulkers} E 2017 {\em \mnras\/} {\bf 464} L6--L10 (\textit{Preprint}
  \eprint{1608.06801})

\bibitem{steiner:2010}
{Steiner} A~W, {Lattimer} J~M and {Brown} E~F 2010 {\em Astrophysical
  Journal\/} {\bf 722} 33--54 (\textit{Preprint} \eprint{1005.0811})

\bibitem{ozel:2010}
{{\"O}zel} F, {Baym} G and {G{\"u}ver} T 2010 {\em \prd\/} {\bf 82} 101301
  (\textit{Preprint} \eprint{1002.3153})

\bibitem{bhattacharyya:2006}
{Bhattacharyya} S and {Strohmayer} T~E 2006  {\bf 636} L121--L124
  (\textit{Preprint} \eprint{arXiv:astro-ph/0509369})

\bibitem{bhattacharyya:2007}
{Bhattacharyya} S and {Strohmayer} T~E 2007  {\bf 666} L85--L88

\bibitem{galloway:2017}
{Galloway} D~K and {Keek} L 2017 {\em ArXiv e-prints\/} (\textit{Preprint}
  \eprint{1712.06227})

\bibitem{SPIT_ETAL02}
{Spitkovsky} A, {Levin} Y and {Ushomirsky} G 2002 {\em Astrophysical Journal\/}
  {\bf 566} 1018--1038

\bibitem{Timmes00}
{Timmes} F~X 2000 {\em \apj\/} {\bf 528} 913--945 source code obtained from
  {\tt http://cococubed.asu.edu/code\_pages/kap.shtml}

\bibitem{woosley-xrb}
{Woosley} S~E, {Heger} A, {Cumming} A, {Hoffman} R~D, {Pruet} J, {Rauscher} T,
  {Fisker} J~L, {Schatz} H, {Brown} B~A and {Wiescher} M 2004 {\em
  Astrophysical Journal Supplement\/} {\bf 151} 75--102

\bibitem{fisker:2008}
{Fisker} J~L, {Schatz} H and {Thielemann} F~K 2008 {\em \apjs\/} {\bf 174}
  261--276

\bibitem{schatz:rp1999}
{Schatz} H, {Bildsten} L, {Cumming} A and {Wiescher} M 1999 {\em Astrophysical
  Journal\/} {\bf 524} 1014--1029 (\textit{Preprint}
  \eprint{arXiv:astro-ph/9905274})

\bibitem{rpprocess}
{Schatz} H, {Aprahamian} A, {Barnard} V, {Bildsten} L, {Cumming} A, {Ouellette}
  M, {Rauscher} T, {Thielemann} F and {Wiescher} M 2001 {\em Physical Review
  Letters\/} {\bf 86} 3471--3474 (\textit{Preprint}
  \eprint{arXiv:astro-ph/0102418})

\bibitem{schatz_rp}
{Schatz} H 2006 {\em International Symposium on Nuclear Astrophysics - Nuclei
  in the Cosmos\/} p 2.1

\bibitem{Lin:2006}
{Lin} D~J, {Bayliss} A and {Taam} R~E 2006 {\em \apj\/} {\bf 653} 545--557

\bibitem{xrb}
{Malone} C~M, {Nonaka} A, {Almgren} A~S, {Bell} J~B and {Zingale} M 2011 {\em
  Astrophysical Journal\/} {\bf 728} 118 (\textit{Preprint} \eprint{1012.0609})

\bibitem{xrb2}
{Malone} C~M, {Zingale} M, {Nonaka} A, {Almgren} A~S and {Bell} J~B 2014 {\em
  Astrophysical Journal\/} {\bf 788} 115

\bibitem{xrb3}
{Zingale} M, {Malone} C~M, {Nonaka} A, {Almgren} A~S and {Bell} J~B 2015 {\em
  Astrophysical Journal\/} {\bf 807} 60 (\textit{Preprint} \eprint{1410.5796})

\bibitem{ABRZ:I}
Almgren A~S, Bell J~B, Rendleman C~A and Zingale M 2006 {\em Astrophysical
  Journal\/} {\bf 637} 922--936 paper I

\bibitem{ZINGALE_ETAL01}
{Zingale} M, {Timmes} F~X, {Fryxell} B, {Lamb} D~Q, {Olson} K, {Calder} A~C,
  {Dursi} L~J, {Ricker} P, {Rosner} R, {MacNeice} P and {Tufo} H~M 2001 {\em
  Astrophysical Journal Supplement\/} {\bf 133} 195--220

\bibitem{harpole:2018}
{Harpole} A and {Hawke} I 2018 {\em ArXiv e-prints\/} (\textit{Preprint}
  \eprint{1806.07301})

\bibitem{cavecchi:2012}
{Cavecchi} Y, {Watts} A~L, {Braithwaite} J and {Levin} Y 2013 {\em Monthly
  Notices of the Royal Astronomical Society\/} {\bf 434} 3526--3541
  (\textit{Preprint} \eprint{1212.2872})

\bibitem{art-2015-cavecchi-etal}
{Cavecchi} Y, {Watts} A~L, {Levin} Y and {Braithwaite} J 2015 {\em \mnras\/}
  {\bf 448} 445--455 (\textit{Preprint} \eprint{1411.2284})

\bibitem{art-2016-cavecchi-etal}
{Cavecchi} Y, {Levin} Y, {Watts} A~L and {Braithwaite} J 2016 {\em \mnras\/}
  {\bf 459} 1259--1275 (\textit{Preprint} \eprint{1509.02497})

\bibitem{art-2017-wilhod-etal}
{Wilson-Hodge} C~A, {Ray} P~S, {Gendreau} K, {Chakrabarty} D, {Feroci} M,
  {Maccarone} T, {Arzoumanian} Z, {Remillard} R~A, {Wood} K, {Griffith} C and
  {STROBE-X Collaboration} 2017 {\em American Astronomical Society Meeting
  Abstracts\/} ({\em American Astronomical Society Meeting Abstracts\/} vol
  229) p 309.04

\bibitem{art-2016-zhang-etal}
{Zhang} S~N {\em et~al.\/} 2016 {\em Space Telescopes and Instrumentation 2016:
  Ultraviolet to Gamma Ray\/} ({\em Proc.~SPIE\/} vol 9905) p 99051Q
  (\textit{Preprint} \eprint{1607.08823})

\bibitem{intZand2018}
in~'t Zand J~J~M {\em et~al.\/} 2018 {\em Science China Physics, Mechanics {\&}
  Astronomy\/} {\bf 62} 29506 ISSN 1869-1927
  \urlprefix\url{https://doi.org/10.1007/s11433-017-9186-1}

\bibitem{castro}
{Almgren} A~S, {Beckner} V~E, {Bell} J~B, {Day} M~S, {Howell} L~H, {Joggerst}
  C~C, {Lijewski} M~J, {Nonaka} A, {Singer} M and {Zingale} M 2010 {\em
  Astrophysical Journal\/} {\bf 715} 1221--1238 (\textit{Preprint}
  \eprint{1005.0114})

\bibitem{astronum:2017}
{Zingale} M, {Almgren} A~S, {Barrios Sazo} M~G, {Beckner} V~E, {Bell} J~B,
  {Friesen} B, {Jacobs} A~M, {Katz} M~P, {Malone} C~M, {Nonaka} A~J, {Willcox}
  D~E and {Zhang} W 2017 {\em ArXiv e-prints\/} Accepted to Proceedings of
  AstroNum 2017 (\textit{Preprint} \eprint{1711.06203})

\bibitem{wdmergerI}
{Katz} M~P, {Zingale} M, {Calder} A~C, {Swesty} F~D, {Almgren} A~S and {Zhang}
  W 2016 {\em Astrophysical Journal\/} {\bf 819} 94 (\textit{Preprint}
  \eprint{1512.06099})

\bibitem{strang:1968}
{Strang} G 1968 {\em {SIAM J. Numerical Analysis}\/} {\bf 5} 506--517

\bibitem{SDC-old}
Nonaka A, Bell J~B, Day M~S, Gilet C, Almgren A~S and Minion M~L 2012 {\em
  Combustion Theory and Modelling\/} {\bf 16} 1053--1088 (\textit{Preprint}
  \eprint{https://doi.org/10.1080/13647830.2012.701019})
  \urlprefix\url{https://doi.org/10.1080/13647830.2012.701019}

\bibitem{vode}
Brown P~N, Byrne G~D and Hindmarsh A~C 1989 {\em SIAM J. Sci. Stat. Comput.\/}
  {\bf 10} 1038--1051

\bibitem{Wang2012190}
Wang W, Shu C~W, Yee H and Sj??green B 2012 {\em Journal of Computational
  Physics\/} {\bf 231} 190 -- 214 ISSN 0021-9991
  \urlprefix\url{http://www.sciencedirect.com/science/article/pii/S0021999111005250}

\bibitem{kappeli:2016}
{K{\"a}ppeli} R and {Mishra} S 2016 {\em \aap\/} {\bf 587} A94

\bibitem{weinan2011principles}
Weinan E 2011 {\em Principles of multiscale modeling\/} (Cambridge University
  Press)

\bibitem{Motheau2018}
Motheau E, Duarte M, Almgren A and Bell J~B 2018 {\em Journal of Computational
  Physics\/} {\bf 372} 1027--1047 ISSN 00219991
  \urlprefix\url{https://linkinghub.elsevier.com/retrieve/pii/S0021999118300469}

\bibitem{Harpole2018}
Harpole A 2018 {\em {Multiscale modelling of neutron star oceans}\/} Ph.D.
  thesis University of Southampton
  \urlprefix\url{https://eprints.soton.ac.uk/id/eprint/422175}

\bibitem{yt}
{Turk} M~J, {Smith} B~D, {Oishi} J~S, {Skory} S, {Skillman} S~W, {Abel} T and
  {Norman} M~L 2011 {\em Astrophysical Journal Supplement\/} {\bf 192} 9
  (\textit{Preprint} \eprint{1011.3514})

\end{thebibliography}

\end{document}